# The Ground State of Graphene and Graphene Disordered by Vacancies

N. Kheirabadi[1]*, A. Shafiekhani[2,3]

[1]Department of Physics, IAU, Northern Tehran Branch, Tehran, 1667934783, Iran.

[2]Physics Department, Alzahra University, Vanak, Tehran, 1993893973, Iran.

[3]School of Physics, Institute for Research in Fundamental Sciences (IPM), P.O.Box:19395-5531, Tehran, Iran.

**Abstract**

Graphene clusters consisting of 24 to 150 carbon atoms and hydrogen termination at the zigzag boundary edges have been studied, as well as clusters disordered by vacancy(s). Density Function Theory and Gaussian03 software were used to calculate graphene relative stability, desorption energy, band gap, density of states, surface shape, dipole momentum and electrical polarization of all clusters by applying the hybrid exchange-correlation functional Beke-Lee-Yang-Parr. Furthermore, infrared frequencies were calculated for two of them. Different basis sets, 6-31g**, 6-31g* and 6-31g, depending on the sizes of clusters are considered to compromise the effect of this selection on the calculated results. We found that relative stability and the gap decreases according to the size increase of the graphene cluster. Mulliken charge variation increase with the size. For about 500 carbon atoms, a zero HOMO-LUMO gap amount is predicted. Vacancy generally reduces the stability and having vacancy affects the stability differently according to the location of vacancies. Surface geometry of each cluster depends on the number of vacancies and their locations. The energy gap changes as with the location of vacancies in each cluster. The dipole momentum is dependent on the location of vacancies with respect to one another. The carbon-carbon length changes according to each covalence band distance from the boundary and vacancies. Two basis sets,

---

*Corresponding author: E-mail address: narjeskheirabadi@gmail.com (N.Kheirabadi)

6-31g* and 6-31g**, predict equal amount for energy, gap and surface structure, but charge distribution results are completely different.

**1. Introduction**

A Graphene sheet is a 2-dimensional honey-comb lattice with carbon atoms occupying the corners of hexagons [1], in which each atom is connected to three other carbon atoms. Due to the observed physical [2], electronic [3] and material science properties [4, 5] of graphene, this material has become an extremely promising and an absolutely vital element for basic sciences and modern technology. The interest in graphene is wide, from quantum electrodynamics [4, 6], condensed matter [7-9] to genomic applications [10, 11].

Several research groups have studied various properties and structures of graphene by different methods, like tight binding [12], semi-empirical [13], molecular dynamics [14-16], Hubbard model [17] and Density Function Theories (DFT) [18, 19].

The use of DFT in the *ab initio* calculation of molecular properties has increased dramatically. This can be attributed to (i) the development of new and more accurate density functions, (ii) the increasing versatility, efficiency, and availability of DFT codes and most importantly, (iii) the superior ratio of accuracy to efforts exhibited by DFT computations relative to other *ab initio* methodologies [20]. As a result, DFT has been accepted as one of the most accurate methods of calculation for a wide range of studies about graphene [21-23]. It is well understood that the edge effect of finite graphene has a crucial role in its electronic properties [24, 25] and DFT method has been used to calculate boundary terminated graphene by hydrogen atom properties [21-23].

On the other hand, the experimental growth of graphene sheets, however, is observed to produce samples disordered or doped by various defects, in which the vacancies are somehow natural in the growth procedure. These defects, occurring due to vacancy formation in finite size graphene, could form scattering centers and affect the carrier mobility and thus adversely affect device performance. Understanding the nature of these defects also assists the ultimate goal of reducing or eliminating defects. DFT calculations have been made to elucidate the nature of graphene defects, but similar to all other quantum calculation methods, calculation complexity increases rapidly by the system size [23], which is a common problem. Furthermore, there were some attempts and calculations to understand the surface shape and magnetism properties of infinite and finite graphene disordered by vacancies [17, 21-23]. The electron structure of one vacancy [21] and two vacancies [17] and voids for ribbon graphene has also been taken into consideration [17].

DFT method with 6-31g* Basis Set (BS) has been used before to study chemical bonds of Quantum Dots (QDs). It was shown that HOMO-LUMO gap decreased with respect to the size increment, and the average Carbon-Carbon (C-C) lengths depended on their proximity to the central ring [17, 21-27]. Furthermore, the properties of a mono-vacancy in infinite graphene sheet with 128 atoms have been studied [28]. In this study, there has been an attempt to extend these studies for more vacancies and their properties.

We focused on a theoretical study of graphene islands with and without vacancy(s). From this point of view, we studied quasi-circular graphene with 24 to 150 carbon atoms, hydrogen at the zigzag boundary edges, as well as in the clusters with one or more vacancy(s). In this article, DFT and Gaussian03 software [29] using the Beke-Lee-Yang-Parr exchange-correlation meta function [30, 31] have been used to study energy, band gap, Density Of States (DOS), surface shape, dipole momentum, and electrical polarization of all planned clusters. In addition, for two of them, the infrared (IR) spectrum is also calculated. Two different BSs, 6-31g* and 6-31g, depending on the size of clusters, have been selected. Moreover, for some cases, 6-31g** calculations have also been used [32, 33].

These research results would help researchers to tailor the properties of graphene QDs. For example, electron gaps between the Highest Occupied Molecular Orbital (HOMO) and the Lowest Unoccupied Molecular Orbital (LUMO) specify metal-semiconductor properties [21]. The relative stability specifies the clusters which show a better performance at higher temperatures. In addition, by extrapolation of data, extra detail of the behavior of graphene cluster properties has been discovered [22].

**2. Graphene Clusters:**

Circular graphene clusters have been considered to pass through 24 to 150 carbon atoms (Fig. 1). By this procedure, there will be four highly symmetric H-terminated clusters: Coronene ($C_{24}H_{12}$), Charcoal ($C_{54}H_{18}$) $C_{96}H_{24}$ and $C_{150}H_{30}$; because of the symmetry, edge effect is important in quantum calculations. These optional clusters are named "knb", where "n" sign hands out the benzene ring numbers around the central benzene ring for each cluster. The next class of clusters is the graphene disordered by vacancy(s). These clusters are named "knbmv"; in this popular nomenclature, "knb" refers to the finite graphenes and it has been abbreviated as it was described earlier (Fig. 1) and "m" specifies the number of vacancy(s) for each cluster.

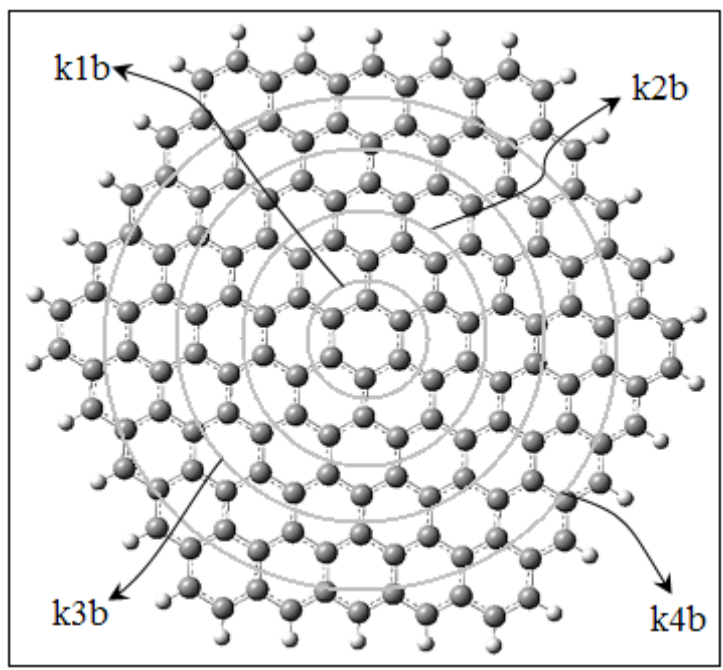

Fig. 1. The name of each circular graphene corresponds to the number of benzene rings around the central benzene ring shown. The Hydrogen at the boundary has been ignored. This nomination is used for abbreviation.

These clusters are: k1b1v, k2b1v, k3b1v, k4b1v, vacant carbon atoms marked by red, k2b2v, k3b2v-1 and k4b2v-1 marked with dark blue, k3b2v-2 and k4b2v-2 marked with yellow, k3b3v-1 and k4b3v marked with green, k4b4v-1 marked with purple and k4b4v-2 marked with light blue, which are shown in Fig. 2. In this case, there are 13 graphene clusters disordered by vacancy(s).

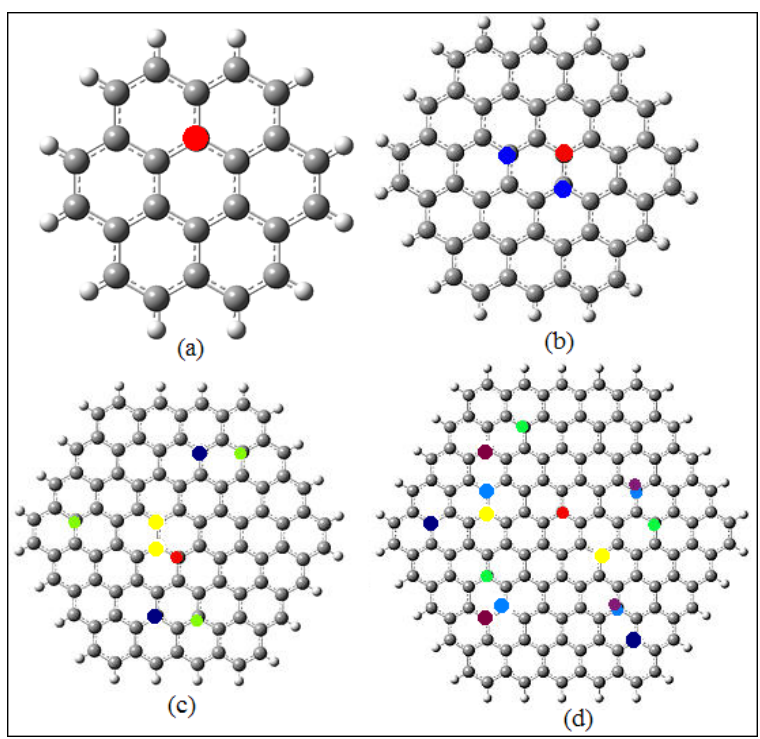

Fig. 2. Configuration of carbon in "knb" clusters with the vacancies is shown for n=1(a), n=2 (b), n=3 (c), and n=4 (d). k1b1v, k2b1v, k3b1v, k4b1v, vacant carbon atoms marked with red, k2b2v, k3b2v-1 and k4b2v-1 marked with dark blue, k3b2v-2 and k4b2v-2 marked with yellow, k3b3v and k4b3v marked with green, k4b4v-1 marked with purple and k4b4v-2 marked with light blue.

## 3. Results

3-1. Graphene

For "knb" group of graphene, for n<4, the 6-31g* and 6-31g** BS and for n=4, 6-31g BS was used [10]. The surface shapes of the clusters, are flat with a deviation of $10^{-5}$Å, and the calculation emphasizes that the energy amount decreases linearly, with the size of graphene (Fig. 3a). In Fig. 3a, it can be seen that the effect of BS does not change energy optimization results with a good approximation; as a result, we can use a smaller BS to calculate large-size graphene structure stable state energy. The relative stabilities of different PAH's often based on the aromatic stabilization energies per carbon. Consequently, according to the Table 1, relative stability decreases according to the size increase of the graphene cluster.
HOMO-LUMO gap decrease logarithmically (Fig. 3b). In a similar way, it can be shown that it does not change gap results either. Based on logarithmic function extrapolation, shown in Fig. 3b, a circular graphene with 500 atoms or more has a gap near to zero. It seems that graphene structure, consisting of 500 carbon atoms, is the threshold of infinite size with zero gap energy.

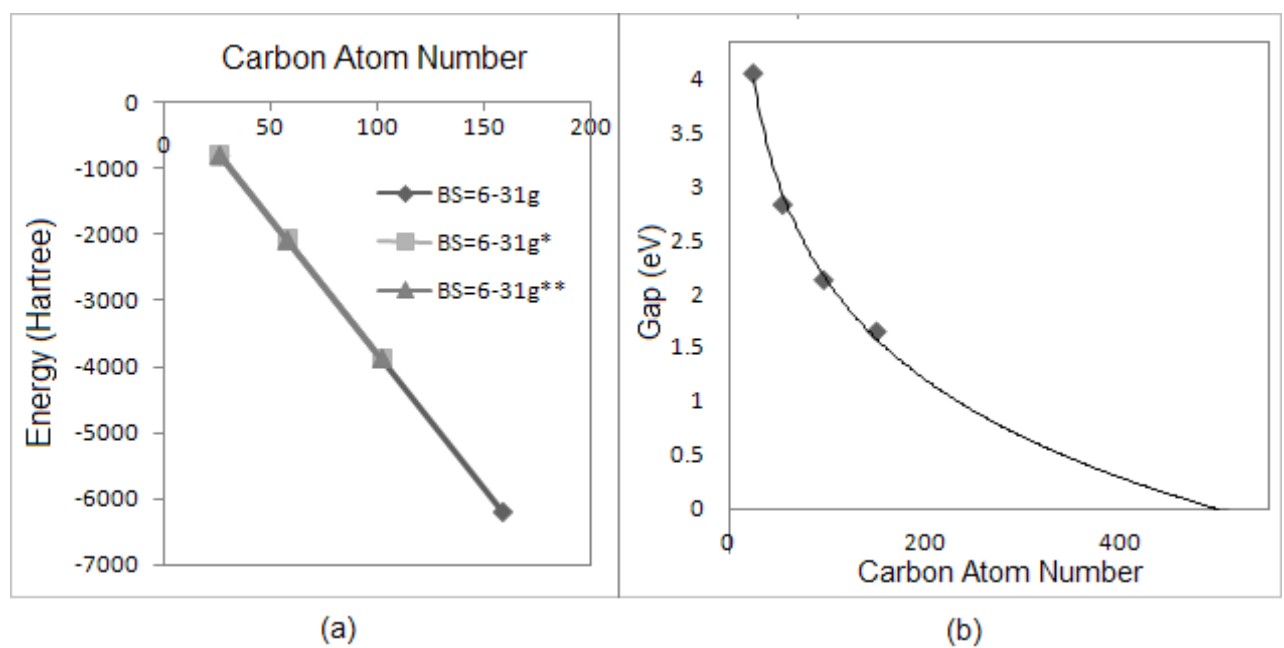

Fig. 3. (a) Indicates that the energy values predicted by the three BSs (6-31g, 6-31g* and 6-31g**) are almost equal and show a linear behavior with the increasing number of carbon atoms. Similar results can also be achieved about cluster gaps. (b) HOMU-LUMO gap energy of "knb" clusters, for n≤4, has been shown by the black circle. The solid logarithmic curve shows the trend of the energy gap going to zero around 500 carbon atoms.

Fermi levels energy increases by the size of the cluster according to the DOS results achieved by Gauss Sum software [34], and the LUMO energy level also decreases. The electron levels become closer, and DOS diagram figure tends to configure electrons like infinite graphene. Graphene electron levels are not spin polarized, and none of the clusters have dipole momentum (Fig. 4).

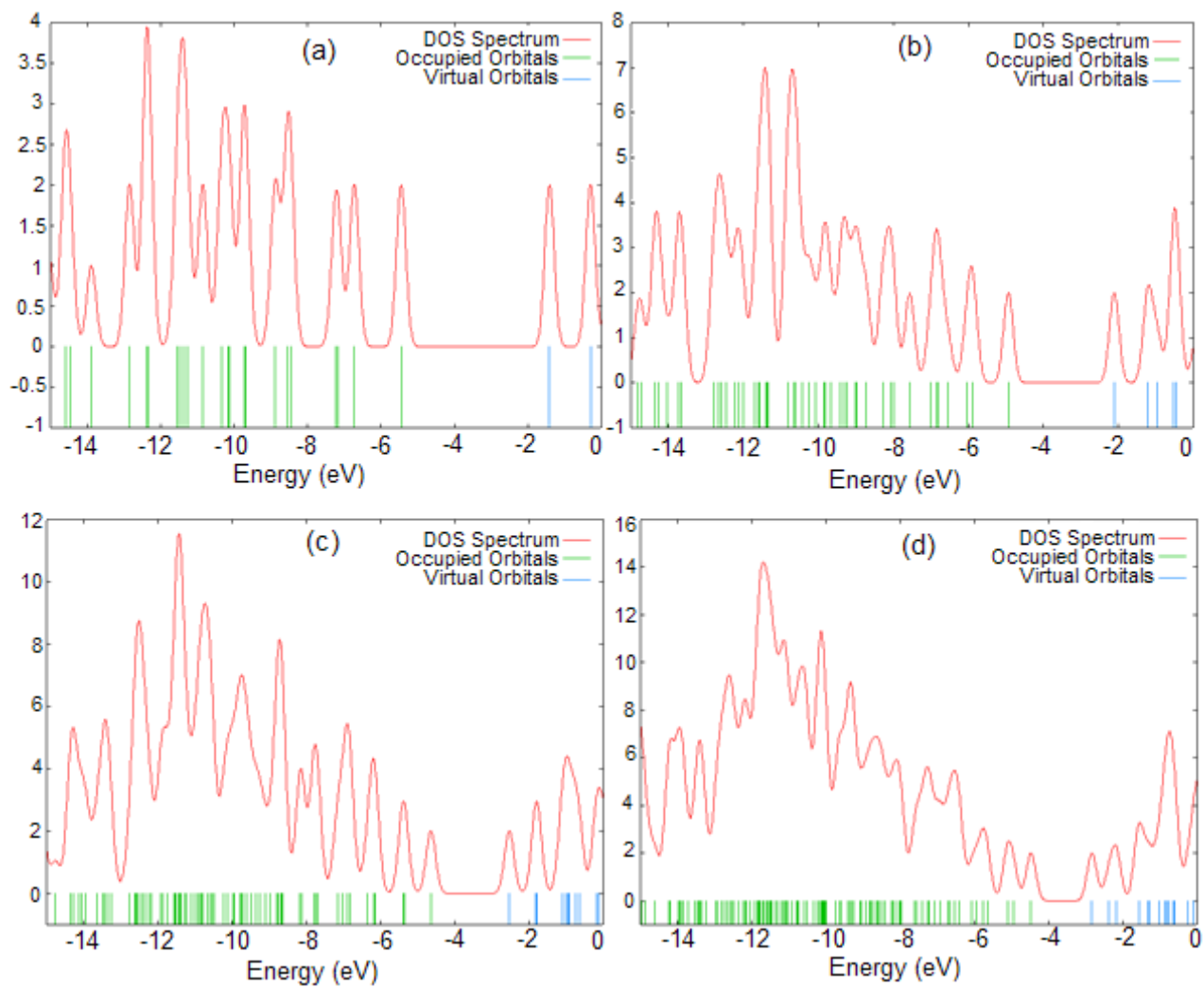

Fig. 4. DOS diagram of (a) k1b, (b) k2b, (c) k3b, and (d) k4b clusters. These figures show that electron levels are not spin polarized. Fermi level energy increases by the size of the cluster, and the LUMO energy level will have the red shift. The electron levels become closer, and DOS diagram figure tends to configure electrons like infinite graphene.

BS is a drastic parameter for clusters charge distribution, so we have avoided comparing the results of charge distribution for different BSs. According to Mulliken charge calculation [32], charge variation intervals of k1b, k2b and k3b increase by their size. As long as an appropriate BS plays a significant role for charge calculations, any comparison between the above clusters and k4b has been avoided [32]. For each four clusters, the charge amount (positive or negative) of each atom increases if that atom is nearer to the edge; also, the hydrogen atoms have equal positive charges.

Table. 1. The summarized results of finite graphene clusters

| Cluster Name | No. of C | BS | Energy (Hartree) | Relative Stability (Hartree) | Gap(eV) HOMO-LUMO | Charge of Atoms (eu) |
|---|---|---|---|---|---|---|
| k1b | 24 | 6-31g* | -921.90 | -38.41 | 4.04 | -0.21 to +0.21 |
| k1b | 24 | 6-31g** | -921.92 | -38.41 | 4.03 | -0.14 to +0.14 |
| k2b | 54 | 6-31g* | -2068.95 | -38.31 | 2.82 | -0.36 to +0.36 |
| k2b | 54 | 6-31g** | -2068.98 | -38.31 | 2.82 | -0.26 to +0.26 |
| k3b | 96 | 6-31g* | -3673.40 | -38.27 | 2.12 | -0.37 to +0.37 |
| k3b | 96 | 6-31g** | -3673.44 | -38.27 | 2.11 | -0.27 to +0.27 |
| k4b | 150 | 6-31g | -5733.93 | -38.23 | 1.64 | -0.26 to +0.26 |

The lengths of C-C bonds depend on their distance from the geometrical center of the cluster. C-C length decreases by the size as known before [22]. K1b cluster's IR frequencies are positive frequencies, and all the diagram peaks happen in a logical fraction of frequencies; for instance, the highest peak indicates truly that there are Carbon-Hydrogen (C-H) bonds in k1b cluster.

3-2. Graphene cluster disordered by vacancy(s)

In order to compare the stability of differently treated samples, desorption energy, $E_{des}$, has been calculated; it reflects the stability of the chemisorption bonds. In the first step, desorption energy for each vacant cluster has been calculated using optimization energy corresponding to the equation (1):

$$C_aH_b \rightarrow C_{(a-n)}H_b + nC \quad (1)$$

In this equation, $C_aH_b$ is the sign of a typical PAH cluster, and n is the number of separated C atom(s). If the optimization energy of the left side is assumed in the Table 1, the optimization energy for $C_{(a-n)}H_b$ is as described in Table 2, and C optimization energy is -37.78 Hartree, then it is possible to calculate desorption energy. So, desorption energy is calculated corresponding to the following equation:

$$\Delta E_{des} = E[C_{(a-n)}H_b] + n\, E[C] - E[C_aH_b], \qquad (2)$$

The results of the above-mentioned calculations have been demonstrated in the fourth column of the Table 2.

According to the fourth column of the Table 2, if the number of separated C increases, cluster desorption energy would also increase. In the cluster k1b1v, desorption energy is +0.48 Hartree, slightly smaller than k1b; so, k1b1v cluster is less stable than k1b, as do the other vacant clusters. IR frequency calculation also proves the optimized k1b1v cluster stability at zero Kelvin temperature.

However, as it is shown in Table 2, the desorption energy of clusters disordered by the same number of vacancies depends on the location of vacancies. For example, both k3b2v-1 and k3b2v-2 clusters have two vacancies in their structures, but there is 0.39 Hartree difference between their desorption energy (Table 2). Also, in case of k4b vacant clusters, for k4b4v-1 and k4b4v-2 clusters, each cluster has four vacancies, but their desorption energy are different. So, there are three important factors affecting the desorption energy of a vacant cluster; firstly, the size. When size varies from k1b through k4b, bigger vacant clusters' desorption energy will increase in comparison to smaller models, in this arrangement. Second, each model's desorption energy depends on the number of vacancies. Then, the stability of clusters with a similar number of vacancies also depends on the vacancy(s) location.

Table. 2. The summarized results of k1b, k2b, k3b and k4b clusters disordered by vacancy(s)

| Cluster Name | No. of C | BS | Energy (Hartree) | Desorption Energy (Hartree) | Gap (eV) HOMO-LUMO | Charge of Atoms (eu) | Dipole Momentum (Debye) |
|---|---|---|---|---|---|---|---|
| k1b1v | 23 | 6-31g* | -883.64 | 0.48 | 3.62 | -0.21 to +0.21 | 0.69 |
| k1b1v | 23 | 6-31g** | -883.66 | _ | 3.63 | -0.18 to +0.18 | 0.69 |
| k2b1v | 53 | 6-31g* | -2030.51 | 0.66 | 1.31 | -0.21 to +0.21 | 1.39 |
| k2b1v | 53 | 6-31g** | -2030.53 | _ | 1.31 | -0.27 to +0.27 | 1.40 |
| k2b2v | 52 | 6-31g* | -1992.21 | 1.17 | 1.73 | -0.33 to +0.33 | 1.95 |
| k3b1v | 95 | 6-31g* | -3634.96 | 0.66 | 1.23 | -0.38 to +0.38 | 0.79 |
| k3b1v | 95 | 6-31g** | -3635.00 | _ | 1.23 | -0.27 to +0.27 | 0.80 |
| k3b2v-1 | 94 | 6-31g* | -3596.52 | 1.31 | 1.29 | -0.37 to +0.37 | 0 |
| k3b2v-1 | 94 | 6-31g** | -3596.56 | _ | 1.29 | -0.27 to +0.27 | 0 |
| k3b2v-2 | 94 | 6-31g* | -3596.92 | 0.92 | 1.34 | -0.39 to +0.39 | 1.41 |
| k3b3v | 93 | 6-31g* | -3558.44 | 1.63 | 1.40 | -0.35 to +0.35 | 2.54 |
| k4b1v | 149 | 6-31g | -5695.50 | 0.66 | 1.02 | -0.27 to +0.27 | 1.36 |
| k4b2v-1 | 148 | 6-31g | -5657.33 | 1.04 | 1.12 | -0.28 to +0.28 | 1.38 |
| k4b2v-2 | 148 | 6-31g | -5657.06 | 1.31 | 0.90 | -0.28 to +0.28 | 2.46 |
| k4b3v | 147 | 6-31g | -5618.73 | 1.86 | 1.18 | -0.29 to +0.29 | 2.24 |
| k4b4v-1 | 146 | 6-31g | -5580.37 | 2.44 | 1.14 | -0.28 to +0.28 | 1.64 |
| k4b4v-2 | 146 | 6-31g | -5580.20 | 2.61 | 1.23 | -0.27 to +0.27 | 3.25 |

Similar results obtained for models with 6-31g* and 6-31g** BS. These two different basis sets predict cluster energies with a good accuracy (Table 2), noting that all of calculated energy amounts are similar in an order of $10^{-2}$ Hartree.

As it can be concluded from Table 2, in the case of k1b1v cluster HOMO-LUMO gap reduction, in comparison with k1b cluster, is 0.42 eV; HOMO-LUMO gap of k2b clusters disordered by vacancy(s) is slightly less than the clusters without vacancy(s); this reduction for k2b1v is 1.51 eV and for k2b2v is 1.09 eV. As for larger clusters, the gap amount depends on vacancy location for the same number of vacancies; so for k3b and k4b vacant clusters, the gap is again dependent on the number of vacancies; moreover, the gap changes by changing vacancy location (Table 2). Also, in this arrangement, the gaps vary according to the cluster number: k4b2v-2, k4b1v, k4b2v-1, k4b4v-1, k4b3v, k4b4v-2 and k3b1v (same gap amount), k3b2v-1, k2b1v, k2b2v, k3b2v-2, k3b3v, k2b2v and k1b1v.

As it can be seen in the arrangement above, a gap generally depends on the size, except for

k3b3v, in which the gap is more than the k2b vacant models' gaps. Hence, vacancies' position is a very important factor for electronic devices gap architecture.

The vacancy changes the DOS, so DOS shape is similar to infinite graphene (levels are packed and sharper), but with less intensity of the diagram as a result of clusters' fewer electrons. As an illustration, DOS diagram of k1b1v is sharper compared to k1b cluster, and its peak has a 0.7 eV red shift. As a result, the reasons for HOMO-LUMO gap decrease, specifically, size increase or vacancy, make graphene DOS shape approach infinite graphene DOS diagram shape, so do DOS diagram of k3b vacant clusters. Moreover, clusters with the same vacancy numbers have different DOS diagrams; for example, the energy state of diagram peak occurs in different energy levels (Fig. 5). Finally, as it can be concluded from Table 2, the results of gap amounts for both 6-31g* and 6-31g** BS, calculation results are consistent with a good accuracy. For example, for k1b1v, the difference between gap amount calculations for two basis sets is 0.01 eV, for k2b1v, k3b1v and k3b2v-1 the difference between clusters gap amounts are zero.

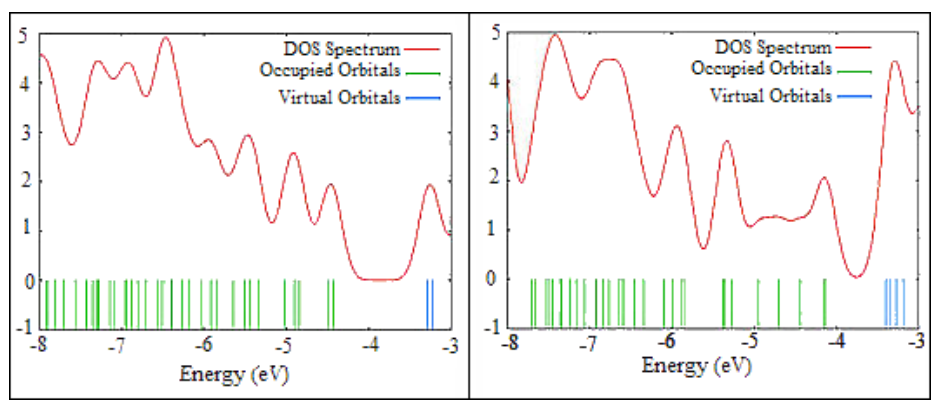

Fig. 5. Right: DOS diagram of k4b2v-1 cluster; Left: DOS diagram of k4b2v-2. As it can be seen, the DOS diagrams of two clusters with the same number of vacancies are different.

For k1b1v cluster, the surface shape is visually warped. The existence of two pentagonal, as a result of vacancy, could be a guide to understand the relation between fullerenes and graphene. Moreover, hydrogen atoms, near the vacancy, pull carbon stronger than others because of their shorter C-H bond length (Fig. 6). For these two bonds, the bond length is 1.0836 Å, while this length for their first neighbor is 1.0855 Å and for next C-H bond, this length is 1.0871 Å (Fig. 6). For 6-31g** BS these results are: 1.0830, 1.0846 and 1.0869 Å, respectively. As a result, the C-H lengths for two different basis sets are so similar, and this result improves the good and acceptably accurate ability of smaller BS for clusters atomic bond length calculations.

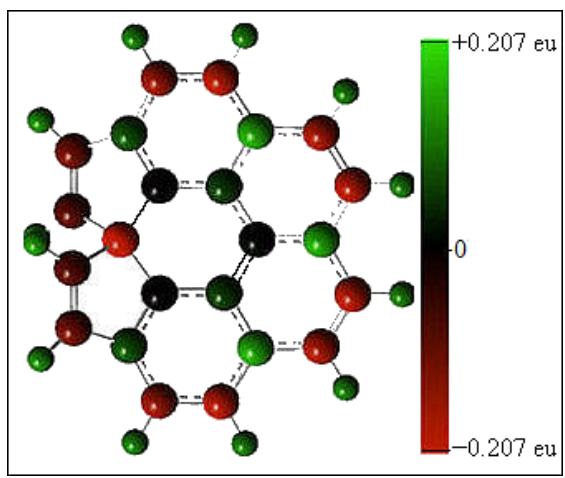

Fig. 6. The charge distribution of the k1b1v cluster based on the basis set of 6-31g**. The effect of vacancy on the bond length and charge distribution between atoms is depicted. Vacancy collapses the structure and electron, so the carbon atom adjacent to it has a higher amount of charge in comparison with other carbon atoms. Similar thing happens to other vacant clusters.

In the case of k2b vacant clusters, the cluster with one vacancy undergoes alteration from planarity, but the cluster with two vacancies undergoes a more significant alteration from planarity and therefore the hexagonal shape of the surface alters as shown in Fig. 7. As shown in Fig. 7, the hexagonal order of the cluster deforms completely, and there are two pentagons, one octagon, and one nonagon in the optimized structure. Meanwhile, the surface shape and the hexagonal order or C-C bond lengths are consistent in the k2b1v cluster. For k2b1v cluster, C-H average length is 1.0872 Å smaller than 1.0876 Å, the average length of k2b C-H bond length, but the average of those C-C bonds adjacent to the vacancy for k2b1v cluster is 1.4002 Å, in comparison to the average of the same C-C bonds in k2b cluster, 1.4225 Å, shows the collapsing property of vacancy (Fig. 7). The k2b1v cluster is flat with $10^{-6}$ Å accuracy[2], but the k2b2v collapses the cluster completely.

The k3b1v cluster collapses due to vacancies, but its surface shape has a hexagonal appearance, while k3b2v-1 is flat with $10^{-6}$ Å accuracy. It seems that the location of the vacancies make two vacancies avoid their effects on surface structure, so do two vacancies for k3b2v-2. K3b3v cluster, which has three vacancies near to its edge, is completely warped. The surface shape of k4b1v is flat; however, it is buckled. In the same way, k4b4v-2 is completely warped. Although, k4b4v-1 cluster has four vacancies; its surface is flat with an accuracy of $10^{-2}$ Å, because four vacancies, symmetrically positioned based on the center, neutralize the effects of one another.

Vacancies nearer to the edge in k4b2v-2, and k4b4v-2 have changed the surface structure dramatically, so that in the last example, there is a nonagon in optimized structure. All in a word, all clusters disordered by vacancy(s), change graphene surface corresponding to the location of vacancy(s) with respect to its boundary, and then vacancies arrangement; similar to the effect of massive balls momentum on a table, stable at its central point.

As a result, for these clusters, C-C lengths are affected by two important factors: the first one is vacancy; it collapses the structure. And the second is hydrogen, it resists against this collapse, and the effect of this factor identifies C-C bond type, because of singlet C-C chemical bonds, which are near the edges, and also near the vacancy portions (Fig. 6 and 7); so, for k4b vacant clusters, different from the last vacant cluster (Table 2), the first three vacant clusters are flat. Therefore, to capture a flat graphene cluster with more vacancies for a special gap reason and 3D architecture, vacancies must be arranged in a symmetric arrangement.

---

[2] Means surface atoms Z parameter is zero with $10^{-6}$ Å accuracy

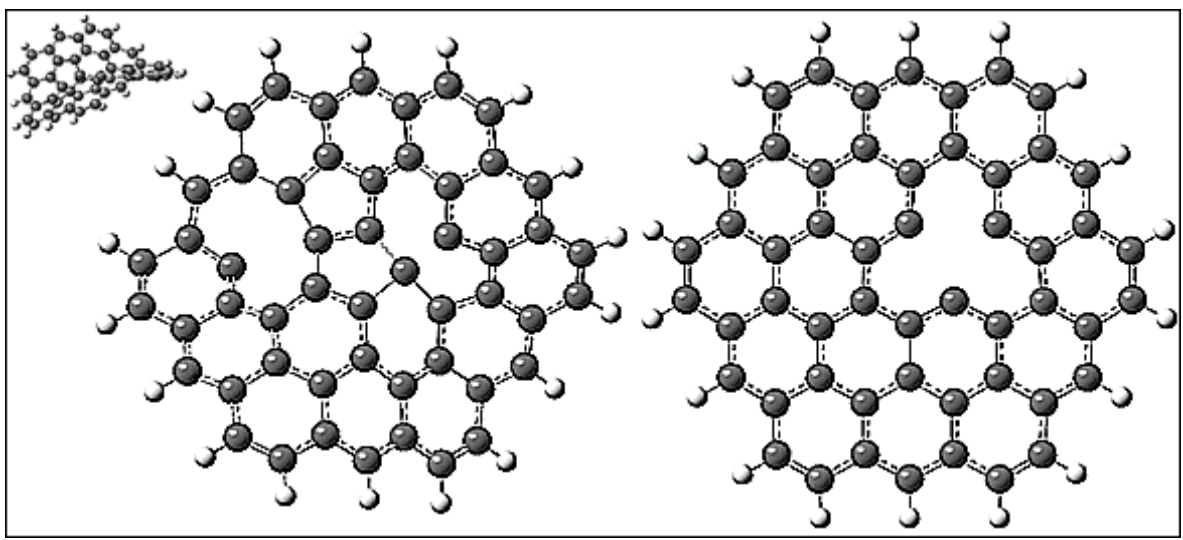

Fig. 7. Configuration of atoms in the k2b cluster shows that the graphene sheet disordered by a vacancy undergoes an alteration from planarity, but the cluster with two vacancies undergoes more significant alterations from planarity. Left figure shows singlet C-C chemical bonds, which are near the edges, and also near the vacancies, as it happens to other larger clusters, insisting on the effect of vacancy and boundary hydrogen atoms which collapse electron and affect C-C bond type.

Atom polarization variation of k1b1v cluster is also smaller than k1b (Table 2). According to Fig. 6, charge of the carbon atoms, adjacent to the vacancy, is strongly negative (-0.16 eu considering an average of 0.02 eu for other carbon atoms), this occurs as a result of electron collapse by vacancy. In the k1b1v cluster, hydrogen atoms, near the vacancy, carry more positive charge compared with other boundary hydrogen atoms (the charge of hydrogen adjacent to the vacancy depending on the distance is: 0.146, 0.137, and 0.131 eu). The vacancy electron density collapses electrons to the vacant positions where electrons are less influenced by the repulsive force; thereby, vacancy warps the cluster (Fig. 6). The k2b1v cluster charge distribution is similar to k2b cluster, (electron polarization is less dense at the center and is denser at edges), while in the k2b2v cluster, the charge distribution has changed dramatically, especially around the vacant center, and this proves that higher number of vacancies results in collapsing the graphene surface more strongly. In this cluster, charge distribution interval is 24 eu above that of k2b1v cluster, which is still less than k2b cluster. This happens as a result of two vacancies existing on the cluster surface. This occurs because of surface warp due to vacancy electron collapse. Surface warp drives back electrons to the edge, so the cluster will be more polarized relative to the stable state flat vacant cluster. This also happens for k3b3v cluster which has a warped surface, and charge collapse centers in the edge, so its charge variation interval is less than all of k3b vacant clusters, and even k3b cluster. Three other k3b vacant clusters are more polarized than k3b cluster; furthermore, all of them have a flat surface. Although k3b2v-1 and k3b2v-2 have fewer numbers of vacancies relative to k3b3v, those are also flat, but the center of collapse is an essential factor. If the collapse center is nearer to the edge, the charge variation interval will also decrease (Tables 1 and 2).

According to the Table 2, in the case of k4b vacant clusters not only vacancy increase charge variation intervals, but also the surface shape affects this variation. This ends in this result that between two clusters with the same number of vacancies, the more warped cluster one will have more polarized atoms. In comparison to non vacant clusters, k1b and k2b vacant clusters' charge interval decreases, meanwhile it increases for k3b and k4b vacant clusters (Tables 2 and 1). According to 6-31g BS route section, charge variation for all k4b vacant clusters are approximately from -0.28 to +0.28 eu.

Considerable results were found for cluster's dipole momentum. Firstly, the k3b2v-1 cluster has zero dipole momentum. This cluster is the one in which vacant carbon atoms are in complete symmetric positions related to central benzene ring. Secondly, the highest dipole momentum amount belongs to k4b4v-2 cluster. Moreover, all vacant clusters, except for k3b2v-1, have dipole momentum (Table 2). So, the symmetry of vacancies relative to central benzene ring is an important factor which affects dipole momentum. In addition, the results of dipole momentum for both 6-31g* and 6-31g** basis sets are the same with a good accuracy (Table 2).

4. Conclusion

The study of quasi-circular graphene culminates in the fact that relative stability and the gap decreases according to the size increase of the graphene segment cluster. Mulliken charge variation increases with the size; moreover, the surface shape is flat. For about 500 carbon atoms, a zero HOMO-LUMO gap amount is predicted. Vacancy reduces the stability and having more than one vacancy affects stability at different rates, according to the location of vacancies. Surface geometry of each cluster depends on the number of vacancies and their locations. The energy gap changes by the location of vacancies in each cluster. The dipole momentum is dependent on the location of vacancies with respect to one another. In addition, the C-C length changes according to each covalence band distance from the boundary and vacancies. The interval of charge variations varies with the number of atoms and surface shape. So, cluster size, vacancy number and surface warp are three factors define atom polarization. Based on the comparison between 6-31g* and 6-31g** BS results, it can be claimed that  energy, gap and surface structure calculations for smaller BS and larger BS are nearly equal with a good accuracy, but the results for the range of charge distribution are completely different.

Acknowledge

The authors would like to thanks Prof. Chandre Dharma-Wardana, Prof. Paul Schleyer, for helpful corrections and reading the manuscript, Dr. Zigerski and Dr. Qaiumzadeh, for their useful discussion. This work was made possible by the facilities of Computational